\documentclass[aps,prd,reprint,superscriptaddress,nofootinbib]{revtex4-1}

\usepackage{amsmath}
\usepackage[dvipsnames]{xcolor}
\usepackage{graphicx}
\usepackage{hyperref}
\usepackage{mathtools}
\usepackage{graphicx}
\usepackage{comment}
\usepackage{cancel}
\usepackage{physics}
\usepackage{mathtools}

\hypersetup{
 bookmarksnumbered=true,
 colorlinks=true,
 linkcolor=[rgb]{0.098,0.098,0.439},
 citecolor=[rgb]{0.098,0.098,0.439},
 urlcolor=[rgb]{0.098,0.098,0.439}
  }

\begin{document}

\title{Dirac-Majorana neutrino type oscillation induced by a wave dark matter}

\author{YeolLin ChoeJo}
\email{particlephysics@kaist.ac.kr}
\affiliation{Department of Physics, KAIST, Daejeon 34141, Korea}
\author{Yechan Kim}
\email{cj7801@kaist.ac.kr}
\affiliation{Department of Physics, KAIST, Daejeon 34141, Korea}
\author{Hye-Sung Lee}
\email{hyesung.lee@kaist.ac.kr}
\affiliation{Department of Physics, KAIST, Daejeon 34141, Korea}

\date{May 2023}

\begin{abstract}
Some properties of a neutrino may differ significantly depending on whether it is Dirac or Majorana type. The type is determined by the relative size of Dirac and Majorana masses, which may vary if they arise from an oscillating scalar dark matter. 
We show that the change can be significant enough to convert the neutrino type between Dirac and Majorana periodically while satisfying constraints on the dark matter.
This neutrino type oscillation predicts periodic modulations in the event rates in various neutrino phenomena including the neutrinoless double beta decay.
As the energy density and, thus, the oscillation amplitude of the dark matter evolves in the cosmic time scale, the neutrino masses change accordingly, which provides an interesting link between the present-time neutrino physics to the early universe cosmology including the leptogenesis.
\end{abstract}

\maketitle

\section{Introduction}

One of the unrevealed properties of the neutrinos is whether they are Dirac type or Majorana type.
Some important physics occur only for the Majorana type: the leptogenesis that can explain the baryon asymmetry of the universe (BAU) and the seesaw mechanism that can explain the smallness of the neutrino masses.
Experiments such as neutrinoless double beta decay can expect signals only for the Majorana neutrinos.

The true nature of the neutrinos may not be simple enough to identify them as either Dirac or Majorana type, though.
It is especially so in view that the properties of the dark matter, which is another mystery in particle physics, are also unrevealed. It is quite possible that neutrino and dark matter are tightly linked, affecting each other.

In this paper, we propose a new scenario in which the neutrino type oscillates due to the wave dark matter.
We show that the amplitude of the oscillating dark matter can be large enough to change the neutrino type back and forth between the Dirac and the Majorana while satisfying all the constraints for dark matter.
It can be tested, for instance, by the modulation search in the neutrinoless double beta decay.
The neutrino type oscillation may not have occurred in the early universe though, allowing the widely-adopted leptogenesis for the BAU.

Coupling an oscillating scalar field to a particle to vary its mass is not new, including the neutrino masses \cite{VanTilburg:2015oza, Arvanitaki:2016fyj, Berlin:2016woy, Zhao:2017wmo, Krnjaic:2017zlz, Brdar:2017kbt, Dev:2020kgz, Huang:2021kam, Dev:2022bae, Huang:2022wmz, Berger:2022tsn, Guo:2022vxr, Davoudiasl:2023uiq, Gherghetta:2023myo}.
Our study differs from the existing works in the sense that it is the first scenario of the neutrino type oscillation.\footnote{We note that it was pointed out in Ref.~\cite{Dev:2022bae}, in which the active-sterile neutrino oscillations in the quasi-Dirac limit were mainly studied, the neutrino might be out of the quasi-Dirac type in the early universe because the Majorana mass term undergoes the suppression over the cosmic time originating from the scaling of the ultra-light dark matter.}

\section{Dirac vs Majorana neutrinos}

The \textit{Dirac neutrino} $\nu_D = \nu_L + \nu_R$ is constructed by left and right-handed Weyl spinors. In contrast, the \textit{Majorana neutrino} $\nu_{M} = \nu_R + \nu_R^c$ is defined using only one Weyl spinor, and satisfies $\nu_{M} = \nu_{M}^c$ condition.
In general, Lagrangian for the neutrino mass for a single flavor can be written as
\begin{equation}
\begin{aligned}
\label{Lagrangian}
\mathcal{L}
= & - m_D \bar{\nu}_D \nu_D 
-\frac{1}{2} m_R \bar{\nu}_{M} \nu_{M}
\\
= & -\frac{1}{2}
\begin{pmatrix}
\bar{\nu}_L & \bar{\nu}_R^c
\end{pmatrix}
\begin{pmatrix}
0 & m_D \\
m_D & m_R
\end{pmatrix}
\begin{pmatrix}
\nu_L^c \\ \nu_R
\end{pmatrix} + h.c.
\end{aligned}
\end{equation}

The light and heavy mass eigenvalues are $
m_{l,h} = \frac{1}{2} ( m_R  \mp \sqrt{ m_R^2 + 4m_D^2 } )$.
The mass matrix is diagonalized in the mass eigenstates $\nu_l$ (light one), $\nu_h$ (heavy one).\footnote{For convenience, we use the following notations interchangeably: $N \equiv \nu_h$, $M \equiv m_h$.}
\begin{align}
\label{massevec}
\begin{pmatrix}
\nu_l \\ \nu_h
\end{pmatrix}
= &
\begin{pmatrix}
\cos \theta_{LR} & \sin \theta_{LR} \\
- \sin \theta_{LR} & \cos \theta_{LR}
\end{pmatrix}
\begin{pmatrix}
\nu_{L}^c \\ \nu_{R}
\end{pmatrix} + h.c.
\\
\text{where } & \sin^2 \theta_{LR} = \frac{1}{2} \left( 1 - \frac{m_R}{\sqrt{m_R^2 + 4m_D^2}} \right) . \nonumber
\end{align}

The ratio of the Majorana and Dirac masses ($m_R / m_D$) determines the left-right (LR) mixing angle $\theta_{LR}$ and the composition of each neutrino mass eigenstate. The $m_R = 0$ case gives a pure Dirac spinor of the degenerate mass eigenvalues $m_D$ and the maximal mixing ($\theta_{LR} = \pi/4$).

For $m_D \gg m_R$ case, the mass eigenvalues are nearly degenerate at $m_D$, and mixing is almost maximal ($\theta_{LR} \simeq \pi/4$).
\begin{align}
\label{QDR} 
|m_{l,h}| & \simeq  m_D (1 \mp 2\delta) ,
\\
\nu_{l,h} & \simeq \pm \frac{1 \pm \delta}{\sqrt{2}}(\nu_L + \nu_L^c) + \frac{1 \mp \delta}{\sqrt{2}} (\nu_R + \nu_R^c) ,
\end{align}
where $\delta = m_R/4m_D$.
This case is called the \textit{quasi-Dirac neutrino} (or pseudo-Dirac neutrino) \cite{Valle:1982yw, Akhmedov:1999uz, Beacom:2003eu, deGouvea:2009fp, Anamiati:2019maf}.
Even though the mass eigenstates are still Majorana spinors, they act similarly to Dirac spinors.
The Dirac mass term becomes more dominant than the Majorana mass term, so the lepton number violation (LNV) process is suppressed in this limit.

In contrast, for the $m_D \ll m_R$ case, the mass eigenvalues are given by the seesaw mechanism \cite{Minkowski:1977sc, Gell-Mann:1979vob, Mohapatra:1979ia, Yanagida:1979as}.
\begin{align}
\label{seesaw}
m_l & \simeq -\frac{m_D^2}{m_R}, \; m_h \simeq m_R , \\
\nu_l & \simeq \nu_L + \nu_L^c, \;
\nu_h \simeq \nu_R + \nu_R^c , 
\end{align}
and the mixing angle $\theta_{LR} \simeq m_D / m_R$ is very small.
The mass eigenstates behave almost like Majorana spinors.

\begin{figure}[tb]
\includegraphics[width=3.4in]{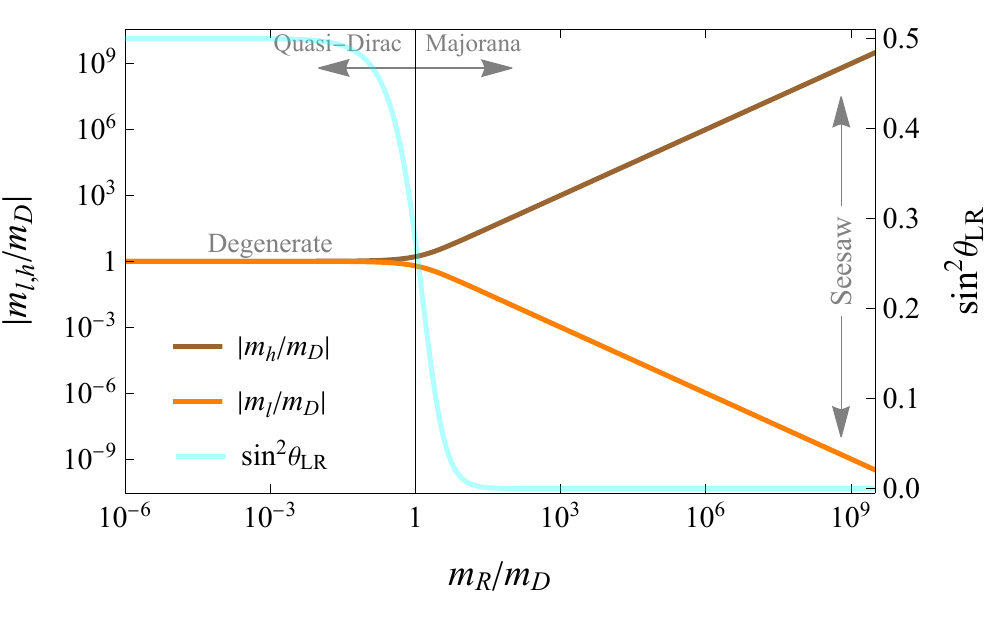}
\caption{\label{evplot} The neutrino mass eigenvalues $|m_{l,h}|$ and the LR mixing angle $\sin^2 \theta_{LR}$ as a function of $m_R/m_D$. In our model, $m_R/m_D$ oscillates widely around $0$ to stay both in quasi-Dirac limit ($m_D \gg m_R$) and Majorana limit ($m_D \ll m_R$).}
\end{figure}

The mass eigenvalues and the LR mixing angle are depicted in Fig.~\ref{evplot} as a function of $m_R/m_D$.

\section{Neutrino type oscillation}

The oscillation between Dirac and Majorana states requires a time variation of $\theta_{LR}$ and thus the variation of $m_R/m_D$.
Such a scenario can be constructed by introducing coupling between neutrino and oscillating scalar field $\phi$ such as a wave dark matter \cite{Hui:2021tkt}, which has periodic behavior for $ 10^{-22}\text{ eV}< m_{\phi} < 30 \text{ eV}$ mass range with corresponding oscillation period is $T = 2\pi / m_{\phi} = \mathcal{O}(\text{ps})\sim\mathcal{O}$(year).
The lower bound originates from the Lyman-$\alpha$ forest data.
The black hole superradiance search may further constrain the mass ranges \cite{Arvanitaki:2009fg, Davoudiasl:2019nlo}.

Assuming the spatial homogeneity of the scalar field, the equation of motion in the expanding universe is
\begin{equation}
\label{EoM}
\ddot{\phi} + 3H \dot{\phi} + \frac{\partial V(\phi)}{\partial \phi} = 0 .
\end{equation}
$H$ is the Hubble parameter and the scalar field potential is $V(\phi) =\frac{1}{2} m_{\phi}^2 \phi^2$.
We assume the mass $m_{\phi}$ remains constant over the cosmic evolution. The Hubble friction is ignored in the current universe ($H_0 \ll m_{\phi}$), so the scalar field has the periodic solution.
\begin{align}
\label{waveDM}
\phi(t) = \phi_0 \cos m_{\phi}t
\end{align}
where $\phi_0 = \sqrt{2\rho_{\phi}} / m_{\phi}$.
The energy density of the oscillating scalar, $\rho_{\phi} = \frac{1}{2} \dot{\phi}^2 + \frac{1}{2}m_{\phi}^2 \phi^2$, has the local density of total dark matter $\rho_{\text{DM}} = 0.3$ GeV/cm$^3$ \cite{ParticleDataGroup:2020ssz} as its upper limit.

There are several ways to introduce interactions between neutrinos and the oscillating scalar field \cite{Berlin:2016woy, Zhao:2017wmo, Krnjaic:2017zlz, Brdar:2017kbt, Dev:2020kgz, Huang:2021kam, Dev:2022bae, Huang:2022wmz, Davoudiasl:2023uiq, Gherghetta:2023myo}.
In general, one can vary both Dirac and Majorana masses by making them functions of an oscillating scalar field, thus periodically changing $m_R/m_D$.
We will consider the case only the Majorana mass is a function of an oscillating scalar field while the Dirac mass $m_D$ is fixed \cite{Zhao:2017wmo, Krnjaic:2017zlz, Brdar:2017kbt, Huang:2021kam, Dev:2022bae, Huang:2022wmz, Davoudiasl:2023uiq}.
Especially, for the definiteness, we take the simplest case where the Majorana mass is given by \cite{Dev:2022bae}
\begin{equation}
\label{interaction}
\mathcal{L} \supset -\frac{1}{2} g\phi \bar{\nu}_R^c \nu_R + h.c.
\end{equation}
This term can serve as an addition to $V(\phi)$ that depends on the density of the right-handed neutrino.
However, since the heavy right-handed neutrinos decay promptly in the early universe, its effect is negligible.
Besides, this coupling generates the effective $\phi^4$ term by the loop with the neutrino \cite{Dev:2022bae, Davoudiasl:2023uiq}, but the desired mass variation behavior of changing $m_R/m_D$ is still guaranteed despite the existence of such a non-linear term \cite{Frasca:2009bc}.

This interaction induces time-varying Majorana mass
\begin{equation}
\label{varymass}
m_R(t) = g \phi(t) \simeq m_{R,0} \cos{m_{\phi}t} ,
\end{equation}
where $m_{R,0} = g \phi_0$ is the current oscillating amplitude of right-handed Majorana mass.\footnote{
In general, there may exist a bare Majorana mass term $-\frac{1}{2} m\bar{\nu}_R^c \nu_R + h.c.$, which makes the center of $m_R(t)$ shifted from zero to $m$.
In this case, the oscillation amplitude $m_{R,0}$ should be similar to or larger than $m$ for the neutrino type oscillation to occur, but we assume $m=0$ here.}
The oscillation amplitude was larger in the early universe because of $\phi_0 = \sqrt{2\rho_{\phi}}/m_{\phi} \propto a^{-3/2}$ behavior where $a$ is the scale factor \cite{Preskill:1982cy, Abbott:1982af, Dine:1982ah}.

In our scenario, the present-time Majorana mass amplitude $m_{R,0}$ is larger than $m_D$, making $m_R(t)$ cross between quasi-Dirac and Majorana regions. We call this phenomenon the \textit{Dirac-Majorana neutrino type oscillation}.
The right-handed Majorana mass term $m_R$ in Eq.~(\ref{Lagrangian}) is now replaced with $m_R(t)$ in Eq.~(\ref{varymass}).
Thus, the $m_R/m_D$ in Fig.~\ref{evplot} may oscillate between the (quasi$\text{-})$Dirac and Majorana limit.
The value of the ratio touches the pure Dirac neutrino case instantly when $m_R = 0$.

\begin{figure}[tb]
\includegraphics[width=3.4in]{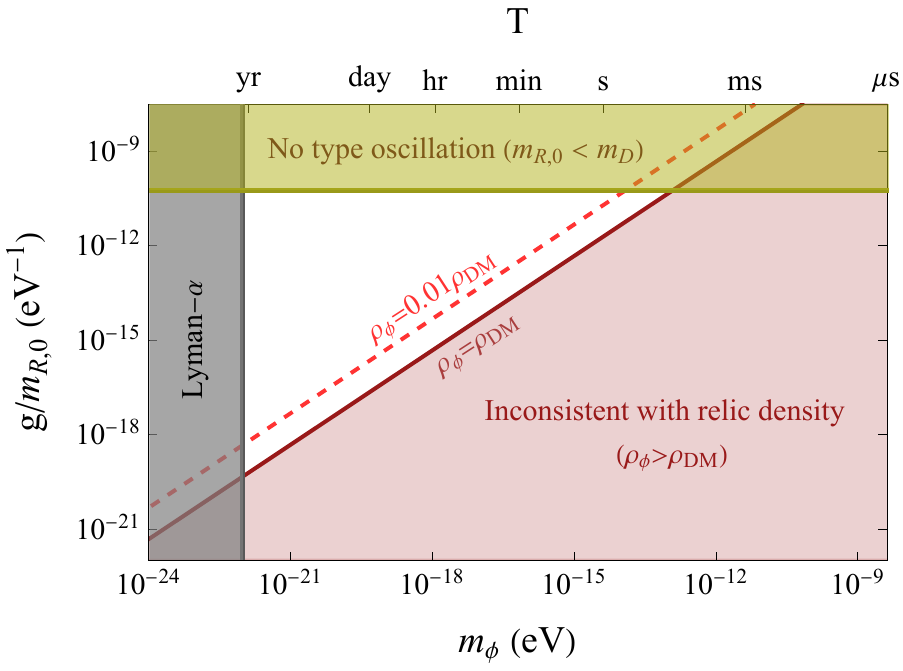}
\caption{
\label{RelicDensity} 
The parameter space of $g/m_{R,0}$ and dark matter mass $m_{\phi}$.
The red region is excluded by the current dark matter relic density. If $m_{R,0} < m_D$, it cannot provide Dirac-Majorana type oscillation and is shaded as the yellow region, with the most conservative assumptions of $g<1$ (perturbative limit) and $m_D = v/\sqrt{2}$ where $v = 246$ GeV is the Higgs vacuum expectation value. For smaller $m_D$ the yellow region will shift upwards.
Lyman-$\alpha$ forest gives the lower bound $10^{-22} \text{ eV} < m_{\phi}$, and the oscillating period $T = \mu$s is the expected time resolution of the neutrinoless double beta decay experiments \cite{Kim:2017xrs}.
Part of the blank region may be sensitive to the black hole superradiance.
A case of the subdominant oscillating dark matter case of $\rho_{\phi} = 0.01\rho_{\text{DM}}$ is drawn with a dashed line for comparison. The density can be as low as $\rho_{\phi} \approx 10^{-18} \rho_{\text{DM}}$, which is the upper-left corner of the blank area.}
\end{figure}

The required coupling strength $g$ to give sufficient amplitude $m_{R,0}$ is given by the following equation.
\begin{equation}
\label{amplitude}
m_{R,0} \simeq 10^{19} \text{ eV} 
\left( \frac{g}{1} \right)
\left( \frac{ \rho_{\phi}}{\rho_{\text{DM}}} \right)^{1/2}
\left( \frac{10^{-22}\text{ eV}}{m_{\phi} } \right) .
\end{equation}
The current dark matter density naturally provides such a large scale of the oscillating Majorana mass with respect to the typical Dirac mass scale of the standard model (SM) fermions, making the oscillating scalar dark matter a good candidate for our scenario. (See Fig.~\ref{RelicDensity}.)

The variation of the neutrino mass may affect neutrino-related physics including the following.
(i) The neutrino flavor oscillation parameters ($\Delta m_{ij}^2$ and $\sin^2 \theta_{ij}$) and the apparent unitarity of the Pontecorvo-Maki-Nakagawa-Sakata (PMNS) matrix for three-flavor mixing of the active neutrinos \cite{Pontecorvo:1957qd, Maki:1962mu}.\footnote{See Refs. \cite{Berlin:2016woy, Krnjaic:2017zlz, Brdar:2017kbt, Dev:2020kgz, Dev:2022bae, Huang:2022wmz, Davoudiasl:2023uiq, Gherghetta:2023myo} for the effect of the mass variation on the neutrino flavor oscillation.}
(ii) The cosmological implications and supernovae/solar neutrino phenomena.
(iii) The lepton flavor violation processes involving the right-handed neutrinos and $W$ boson decay ($W \rightarrow \ell N$).

\begin{figure}[b]
\includegraphics[width=3.4in]{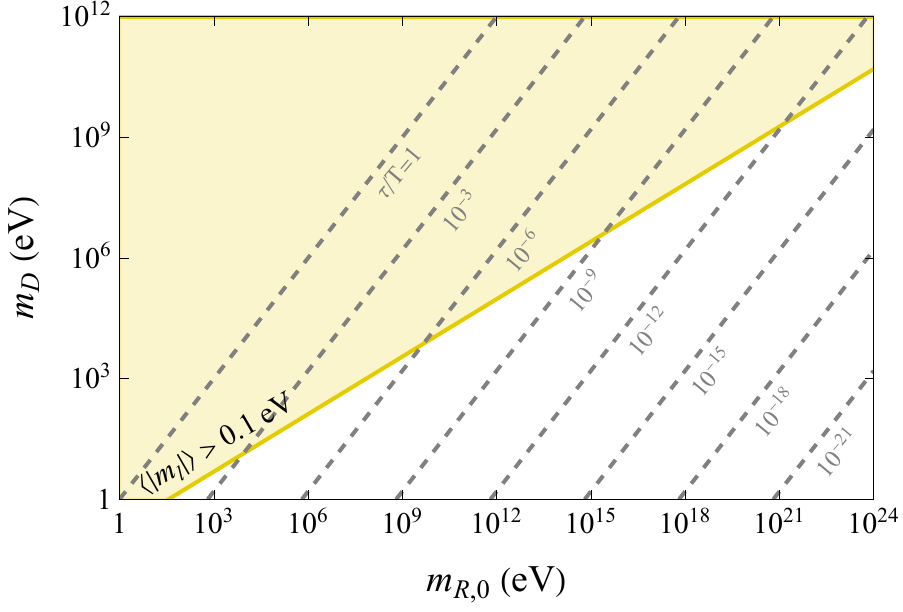}
\caption{
\label{bound}
The parameter space of Majorana mass amplitude $m_{R,0}$ and Dirac mass $m_{D}$. The yellow region is disfavored as the time-averaged light neutrino mass $\expval{\abs{m_{l}}} = \frac{1}{T} \int_0^T \; \abs{m_{l}(t)} dt$ is greater than 0.1 eV, which is about the current constraint on the active neutrino mass.
The dashed lines represent the values of the quasi-Dirac time ratio $\tau/T$.
Only the region with a small $\tau/T$ survives.
}
\end{figure}

Although it is pending the scrutiny of the data with dedicated analysis, we assume it is okay if the time-averaged quantities satisfy the known bounds.
For instance, we impose the active neutrino mass bound \cite{KamLAND-Zen:2016pfg, ParticleDataGroup:2020ssz, Huang:2019tdh, Planck:2018vyg, Majorana:2022udl} on the time-averaged neutrino mass: $\expval{\abs{m_{l}}} = \frac{1}{T} \int_0^T \; \abs{m_{l}(t)} dt < $ 0.1 eV.
The allowed parameter space of Dirac mass $m_{D}$ and Majorana mass amplitude $m_{R,0}$ is shown in Fig.~\ref{bound}.
This time-average analysis can also be applied to the effect of the mass variation on the neutrino flavor oscillation parameters, but it is out of the scope of this paper.

For the convenient analysis, we call $m_{D} > m_{R}(t)$ part the quasi-Dirac region.
The time interval $\tau$ for this region is given by
\begin{align}
\label{taurange}
\frac{m_{D}}{m_{R,0}} = \sin \left(\frac{\pi}{2} \frac{\tau}{T} \right)
\; \text{ for } m_{D} < m_{R,0} ,
\end{align}
and the `quasi-Dirac time ratio' $\tau/T$ is $1$ for $m_{D} > m_{R,0}$ case, meaning the neutrino is continually in the quasi-Dirac region.

\section{Implications on neutrinoless double beta decays}

The neutrinoless double beta decay ($0\nu\beta\beta$) $2n \rightarrow 2p + 2e^- $ as a LNV process \cite{Schechter:1981bd} is one of the best examples that give a distinction for quasi-Dirac and Majorana neutrinos.

The $0\nu\beta\beta$ decay rate is proportional to the square of the following amplitude $\mathcal{A}$.
\begin{align}
\label{effmass}
\mathcal{A} & = \abs{ \sum_i U^{2}_{ei} m_i \mathcal{M}(m_i)} , 
\end{align}
where $U$ is the extended PMNS matrix including LR mixing, and $\mathcal{M}$ is the nuclear matrix element \cite{Doi:1985dx, Rodejohann:2011mu, Faessler:2014kka, Bolton:2019pcu, Dekens:2023iyc}.

For a simple case of a single flavor, the amplitude is
\begin{equation}
\label{two}
\mathcal{A} = \abs{ m_l \cos^2 \theta_{LR} \mathcal{M}(m_{l}) + m_h \sin^2 \theta_{LR} \mathcal{M}(m_{h}) } .
\end{equation}
For pure Dirac neutrinos, the amplitude is exactly zero because the two terms in $\mathcal{A}$ cancel each other precisely.
In the quasi-Dirac region, the amplitude $\mathcal{A}$ is still highly suppressed \cite{Doi:1985dx, Akhmedov:1999uz}.
In the Majorana region, $\mathcal{A}$ is relatively large and the $0\nu\beta\beta$ process is activated.
If the Dirac-Majorana neutrino type oscillation occurs, the $0\nu\beta\beta$ process turns on and off periodically as $m_R(t)$ oscillates.
The three-flavor case shows similar behavior.

The time-averaged amplitude $\expval{\mathcal{A}}$ can be constrained by the current bound from the $0\nu\beta\beta$ experiments \cite{KamLAND-Zen:2016pfg, Majorana:2022udl}.
The Majorana region occurs most of the time according to Fig.~\ref{bound}, in which the LR mixing angle is negligibly small.
Thus, only the first term in Eq.~(\ref{two}) remains, and the $0\nu\beta\beta$ bound is reduced to $\expval{|m_l|}$ bound.

The neutrino type oscillation can be tested by analyzing $0\nu\beta\beta$ data reflecting time modulation. The event number will show a sharp decrease when the state enters the quasi-Dirac region. The time length of the $0\nu\beta\beta$ rate reduction depends on the dark matter mass $m_\phi$ and the quasi-Dirac time ratio $\tau/T$.

Such a modulation search can apply to others, too. 
For instance, the cosmic neutrino background signal rate is larger for the Majorana neutrino compared to the Dirac neutrino by a factor of two \cite{Long:2014zva, PTOLEMY:2019hkd, Hernandez-Molinero:2022zoo}.
A modulation search could be potentially much more sensitive than a typical signal-over-background search for these extremely weak signals.

\section{Implications on leptogenesis}

The SM does not provide an explanation for the observed BAU. Leptogenesis is a possible model for baryogenesis by generating nonzero $B-L$ via the LNV process, in the presence of the Majorana mass term. A decay of a heavy Majorana neutrino generates lepton asymmetry and this nonzero lepton number leads to baryogenesis via the sphaleron process.

The LNV comes from the $CP$ violating decay of heavy Majorana neutrinos $N_i$. The $CP$ asymmetry $\epsilon_i$ for this process is defined as
\begin{align}
\epsilon_i &= \frac{\Gamma(N_i \rightarrow \ell \Phi ) - \Gamma(N_i \rightarrow \bar{\ell} \Phi^{\dagger}) }{\Gamma(N_i \rightarrow \ell \Phi ) + \Gamma(N_i \rightarrow \bar{\ell} \Phi^{\dagger})} ,
\end{align}
where $\ell$ is the lepton doublet and $\Phi$ is the Higgs doublet. To produce enough lepton number to explain the observed BAU, at least $\epsilon_i \approx 10^{-6}$ is needed \cite{Davidson:2008bu}.

The $CP$ asymmetry $\epsilon_i$ is proportional to
\begin{align}
\epsilon_i \propto \sum_{k\neq i} \frac{\text{Im}(y^{\dagger}y)_{ik}^2}{(y^{\dagger}y)_{ii}(y^{\dagger}y)_{kk}} \frac{ (M_i^2 - M_k^2) M_i \Gamma_k}{(M_i^2 - M_k^2)^2 + M_i^2 \Gamma_k^2} ,
\end{align}
where $y$ is the Yukawa matrix for the Dirac mass term and $\Gamma_k$ is the decay rate of $k$-th heavy neutrino.
In the minimal leptogenesis scenario, where very heavy neutrinos are assumed (with a hierarchy $M_1 \ll M_{2,3}$), it reduces to $\epsilon_1 \propto M_1$.
Here, the lower bound $M_1 > 10^{9}$ GeV is required \cite{Davidson:2002qv, Davidson:2008bu}.
In another scenario, called the resonant leptogenesis, nearly degenerate $M_i$'s are assumed, and the mass bound can be as low as TeV scale \cite{Pilaftsis:1997jf, Pilaftsis:2003gt, BhupalDev:2014pfm}.

\begin{figure}[tb]
\includegraphics[width=3.4in]{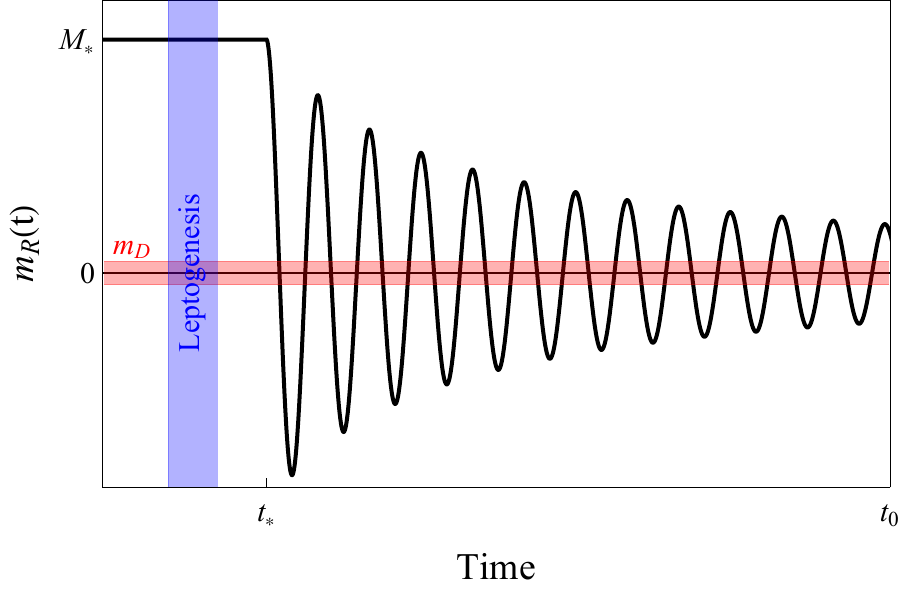}
\caption{
\label{OscDecay}
Schematic illustration of the Majorana mass term over cosmic time. ($t_0 $ is the present time.) The Majorana mass $M_*$ is frozen before $t_*$, at which the oscillation began ($H \approx m_{\phi}$), and the leptogenesis may occur before $t_*$.}
\end{figure}

Since the amplitude of the oscillating scalar field decays as $\phi_0 \propto a^{-3/2}$, the Majorana masses of neutrinos are expected to be far larger in the early universe than the present values when they are functions of the oscillating scalar dark matter \cite{Zhao:2017wmo, Krnjaic:2017zlz, Brdar:2017kbt, Dev:2020kgz, Huang:2021kam, Dev:2022bae, Huang:2022wmz, Davoudiasl:2023uiq}.\footnote{The scalar field becomes radiation-like and $\phi_0 \propto a^{-1}$ if the $\phi^4$ term is dominant \cite{Dev:2022bae, Turner:1983he}, but we set the small $\phi^4$ term here.}
The seesaw mechanism engages in this limit, and $M \simeq m_R$.

The Majorana mass could be considered as a constant before the scalar field oscillation began when the Hubble parameter $H \simeq T^2/M_{\text{Pl}}$ is about the dark matter mass $m_{\phi}$, where $T$ is the temperature of the universe and $M_{\text{Pl}}$ is the Planck mass.
Since the scale factor value $a_*$ at $H \approx m_{\phi}$ can be determined when $m_\phi$ is given, the temperature $T_*$ at the dark matter production stage is given by
\begin{align}
T_* \simeq 10^{3} \text{ eV} \left( \frac{m_{\phi} }{10^{-22}\text{ eV}} \right)^{1/2} .
\end{align}
Using $T \propto 1/a$, $m_R \propto a^{-3/2}$, and $T_0 \simeq 2.73 \text{ K}$, the Majorana mass $M_{*}$ at that stage can be obtained.
\begin{align}
\label{mRratio}
\frac{M_{*}}{M_{0}} \simeq 10^{11} \left( \frac{m_{\phi} }{10^{-22}\text{ eV}} \right)^{3/4} , 
\end{align}
where $M_{0}$ is the current Majorana mass amplitude.

The grand unification scale is typically taken for the heavy Majorana neutrino mass in the minimal leptogenesis, but it can decrease drastically over time.
For $m_\phi=10^{-22}$ eV as a benchmark point, $M_{*} = 10^{15}$ GeV decreases to $M_{0} \sim 10$ TeV at the present time.
This small $M_0$ suppresses $y \propto \sqrt{M_0}$ according to the Casas-Ibarra parametrization \cite{Casas:2001sr}, which is widely adopted to ensure consistency with the neutrino oscillation data. Nevertheless, sufficient $CP$ asymmetry $\epsilon$ can be achieved if the masses are degenerate.

We note that large $M_{*}$ and small $y \propto \sqrt{M_{0}}$ may not provide the thermal production of the heavy Majorana neutrinos in the early universe, but the leptogenesis can be still achieved if, for instance, the thermal equilibrium is obtained with a new gauge boson that couples to the right-handed neutrinos and other particles.
Also, nonthermal production of the Majorana neutrinos from inflaton decays may occur.

The scale of $M_{*}$ for the leptogenesis links with $M_{0}$, which determines the Dirac-Majorana neutrino type oscillation in the present universe. This relation is schematically illustrated in Fig.~\ref{OscDecay}.

\section{Summary and outlook}
We studied the scenario in which the neutrino masses are given as a function of the wave scalar dark matter in the way it results in Dirac-Majorana neutrino type oscillations.
We showed this scenario is consistent with the various constraints on dark matter, and even if the oscillating scalar field is a fraction of the relic dark matter, the scenario works well.

The amplitude of the oscillating dark matter decreases over time, but the leptogenesis may still occur with the frozen Majorana mass in the early universe, while the suppressed oscillating mass amplitude at the present time may result in unique periodic conversion signatures in various neutrino experiments including the neutrinoless double beta decays.
Rich physics and cosmology with neutrino type oscillation are warranted.

\begin{acknowledgments}
We thank Kazuki Enomoto, Jiheon Lee, and Jaeok Yi for their helpful discussions.
We also thank Peter Denton for a useful conversation about the cosmic neutrino background.
This work was partly supported by the National Research Foundation of Korea (Grant No. NRF-2021R1A2C2009718).
\end{acknowledgments}

\bibliography{ref}

\end{document}